\documentclass[english,aps,prb,preprint,onecolumn]{revtex4-1}
\usepackage{verbatim}
\usepackage{amstext}
\usepackage{graphicx}
\makeatletter

\draft % marks overfull lines with a black rule on the right
\usepackage{subfigure}
\usepackage{multirow}
\usepackage{rotating}%\usepackage{breqn}

\newcommand{\titanate}{Li$_2$TiO$_3$}

\newcommand{\etal}{\emph{et al.}}

\makeatother

\usepackage{babel}
\begin{document}

\title{Anisotropic charge screening and supercell size convergence of defect
formation energies}

\author{Samuel T. Murphy}

\email{samuel.murphy05@ic.ac.uk}

\selectlanguage{english}%

\author{Nicholas D. M. Hine}

%\author{Robin W. Grimes}

\affiliation{Department of Materials, Imperial College London, South
Kensington, London, SW7 2AZ, UK.}

\date{\today}
\begin{abstract}
One of the main sources of error associated with the calculation of
defect formation energies using plane-wave Density Functional Theory
(DFT) is finite size error resulting from the use of relatively small
simulation cells and periodic boundary conditions. Most widely-used
methods for correcting this error, such as that of Makov and Payne,
assume that the dielectric response of the material is isotropic and
can be described using a scalar dielectric constant $\epsilon$. However,
this is strictly only valid for cubic crystals, and cannot work in
highly-anisotropic cases. Here we introduce a variation of the technique
of extrapolation based on the Madelung potential, that allows the
calculation of well converged dilute limit defect formation energies
in non-cubic systems with highly anisotropic dielectric properties.
As an example of the implementation of this technique we study a selection
of defects in the ceramic oxide \titanate\ which is currently being
considered as a lithium battery material and a breeder material for
fusion reactors.
\end{abstract}
\maketitle

\section{Introduction}

Point defects play an essential role in a number of important materials
properties such the accommodation of nonstoichiomety and facilitation
of diffusion through a crystal matrix. The difficulties associated with
making direct observations on such small length scales mean it
is desirable for first principles methods such as Density Functional
Theory (DFT) to provide insight into the properties and behaviour
of both intrinsic and extrinsic point defects.

In DFT, point defects are normally modelled using the supercell methodology,
whereby vacancy, interstitial or substitutional defects are placed
in a simulation supercell which is then tesselated though space using
periodic boundary conditions (PBCs) to create an infinite crystal.
Therefore, any defect included in the original supercell will also
be tesselated and the interaction of these defect images can have
a significant influence on the defect formation energy. This problem
is particulary acute in the case of charged defects as the Coulomb
interaction decays slowly as a function of the separation between
point charges\cite{nieminen:convergence}. A number of correction schemes
have been devised to extract the formation energies in the desired dilute
limit from simulations of relatively small supercells: these have been
widely applied to systems such as silicon
\cite{Taylor:coulombic_correction,corsetti:convergence},
NaCl\cite{Taylor:coulombic_correction,schultz:coulombic_correction},
diamond \cite{freysoldt:coulombic_correction,freysoldt:correction2},
GaAs\cite{lany:coulombic_correction,lany:charge_correction,
freysoldt:coulombic_correction,freysoldt:correction2},
InP \cite{castelton:defects_convergence_InP} and
Al$_{2}$O$_{3}$\cite{hine:coulombic_correction,hine:onetep_al2o3}.
Inherent in all of these schemes is the assumption that the dielectric
response of the material is isotropic and can be described by a single
dielectric constant, $\epsilon$. Strictly, this only holds for cubic
systems, but in many cases the degree of anisotropy is modest enough
that the assumption of an isotropic dielectric response is adequate
\cite{hine:coulombic_correction,lany:charge_correction,
malone:defects_in_SiXII}. Intuitively, one might expect that this would
not be the case for many of the more complex crystals that are currently
being proposed for industrial applications, particularly those with layered
structures.

\begin{figure}[P]
\centering{}\includegraphics[width=0.8\columnwidth]{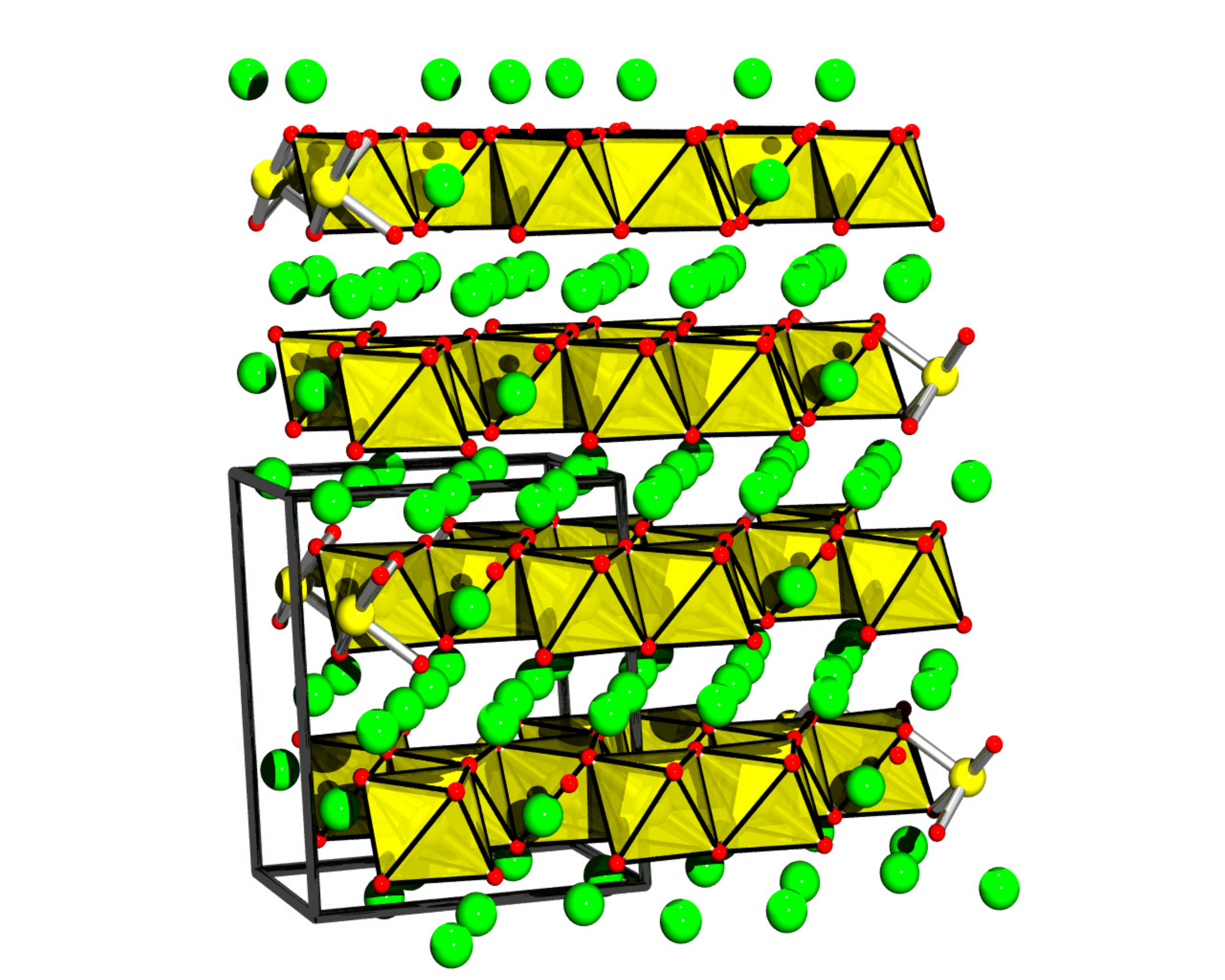}
\caption{\label{figure:supercell}Crystal structure of $\beta-$\titanate.
Yellow, green and red spheres represent titanium, lithium and oxygen
ions respectively. The black outline represents a single unit cell.}\newpage
~
\newpage
\end{figure}

One such system is lithium metatitanate, $\beta$-\titanate, which is
currently under consideration for use in lithium ion
batteries \cite{Zhang:lithiumionbattery} and as breeder material in
fusion reactors \cite{raffray:breederblankets}. \titanate\ may be
described as a distorted rocksalt structure (space group $C2/c$)
with alternating Li, LiTi$_{2}$ and O
planes\cite{lang:Li2TiO3_structure,kataoka:Li2TiO3_structure,murphy:li2tio3}
which are clearly visible in Fig. \ref{figure:supercell}. Within the LiTi$_2$
layers the Ti atoms form a honeycomb structure with a Li ion at the centre of
each hexagon. It is this layered structure that gives rise to the material's
interesting dielectric properties. Currently, not much is known about the
properties of the defects in Li$_2$TiO$_3$. Vijayakumar \emph{et al.}
determined, using empirical potentials, the relative energies required to
remove the different Li atoms and found that the formation energy of a Li
vacancy defect in the pure Li layer is 0.30 eV greater than in the LiTi$_2$
layer\cite{Vijayakumar:Li2TiO3_li_diff}. The linear ``muffin-tin'' orbitals
method has been used to study H substitution onto Li sites where the hydrogen
is observed to move from the lithium site and bond to an oxygen forming a
hydroxide\cite{Zainullina:lihtio3}. One of the principle reasons for this
shortage of theoretical results is the very same problem we try to address
in this paper: namely that the anisotropy of the system means that it is
hard to extract well-converged formation energies.

In this study we investigate the convergence of the formation energies of
point defects in monoclinic $\beta-$\titanate\ as a function of supercell
size. Specifically, we study the V$_{\mathrm{Ti}}^{-4}$,
Li$_{\mathrm{Ti}}^{-3}$ and O$_{i}^{-2}$ defects (modified Kr\"{o}ger-Vink
notation). These defects represent a range of different defect types and also
have high charge states and so are subject to the largest finite-size errors.

\section{Methodology}

The DFT simulations presented here were performed using the plane-wave
pseudo-potential code CASTEP\cite{clark:CASTEP}.
Exchange-correlation is described using the generalised gradient approximation
of Perdew, Burke and Ernzerhof (GGA-PBE)\cite{perdew:gga}. A $\Gamma$-centered
Monkhorst-Pack \cite{Monkhorst:k-pointgris} scheme was used to sample the
Brillouin zone with the separation of points maintained as close as possible
to 0.05 \AA{}$^{-1}$ along each axis.

The same pseudopotentials as in previous work\cite{murphy:li2tio3}
(ultrasoft pseudo-potentials (USPs), generated ``on-the-fly'' in
CASTEP, and normconserving pseudopotentials (NCPPs) from the standard
library in Materials Studio) were employed here. The planewave kinetic
energies were truncated at 550 eV and 1700 eV for the USP and NCPPs
respectively. The Fourier transform grid for the electron density
is larger than that of the wavefunctions by a scaling factor of 2.0
and the corresponding scaling for the augmentation densities was set
to 2.3 when USPs were in use. These values were determined by performing
convergence tests of the energy from self consistent single point
simulations. The lattice parameters determined using DFT are within
1$\%$ of the experimental as shown in Table \ref{table:lattice_parameters}.
Additionally a comparison of all the atomic positions has been included in
the Supplementary Materials.

\begin{table}
	\caption{\label{table:lattice_parameters}Table showing the lattice parameters
and bandgaps calculated using the Ultrasoft and Norm-conserving
pseudo-potentials compared to the available experimental data.}
	\begin{center}
	\begin{tabular}{| c | c | c | c | }
		\hline
	 	Property & Ultrasoft & Norm-conserving & Experimental \\
		\hline
		Volume /\AA$^3$ & 432.98 & 441.53 & 427.01\cite{kataoka:Li2TiO3_structure} \\
		\textit{a} /\AA & 5.09 & 5.12 & 5.06\cite{kataoka:Li2TiO3_structure} \\
		\textit{b} /\AA & 8.83 & 8.90 & 8.79\cite{kataoka:Li2TiO3_structure} \\
		\textit{c} /\AA & 9.80 & 9.85 & 9.75\cite{kataoka:Li2TiO3_structure} \\
        $\beta$ /$^{\circ}$ & 100.19 & 100.24 & 100.21
        \cite{kataoka:Li2TiO3_structure}  \\
        $E_g$ /eV & 3.27 & 3.43 & 3.90\cite{hosogi:li2tio3_bandgap} \\
        \hline
	\end{tabular}
	\end{center}
\end{table}

Supercells for the defect simulations were then constructed by creating
$l\times{m}\times{n}$ repetitions of the relaxed unitcell in $a$, $b$
and $c$ respectively with the resulting supercells containing between 192 and
576 atoms. The lattice parameters and cell angles were fixed during
minimization of the defect containing supercells so only atom positions
were relaxed until the residual force on each atom was $<0.08$ eV
A$^{-1}$ and the difference in energy between consecutive ionic relaxation
steps was $<5\times10^{-5}$ eV atom$^{-1}$.

The calculated band gap of \titanate\ was 3.27 eV for the USPs and
3.43 eV for the NCPPs compared to an experimental value of
3.9 eV\cite{hosogi:li2tio3_bandgap} and values of 2.5-3.5 eV from previous
first principles studies \cite{Zainullina:lihtio3,shein:lithium_mettalates,
wan:li2tio3_DFT}. Underestimation of the bandgap is a common feature of LDA
and GGA calculations: fortunately the lack of occupied states in the band
gap for the fully charged defects investigated here ensures that the
defect formation energies reported are not directly affected by this
source of error\cite{hine:coulombic_correction}, although there are
still effects due to the localization of states. The implementation of
a hybrid functional, such as the HSE functional\cite{heyd:HSE_functional}
would most likely change the defect formation energies. However, it is
important to note that whatever functional is used, formation energies
are still subject to similar finite-size effects, as these are determined
almost entirely by the Coulombic interaction of periodic images, with
functional-dependent effects being limited to the polarizability and
localisation of charges.

Following the formalism of Zhang and Northup \cite{zhang:formationenergy}
the formation energy of a defect is given by:
\begin{equation}
E_{f}=E_{\textrm{defect}}^{T}-E_{\textrm{perf}}^{T}+\sum_{i}{n_{i}}
\mu_{i}+qE_{F}\label{eq:defectformationenergy}
\end{equation}
where $E_{\textrm{defect}}^{T}$ and $E_{\textrm{perf}}^{T}$ are
the DFT total energies of a system with and without the defect, $n_{i}$
is the number of atoms added/removed, $\mu_{i}$ is the chemical potential
of species $i$, $q$ is the charge on the defect and $E_{F}$ is
the Fermi energy (defined here as the valence band maximum, VBM).

Representative chemical potential potentials ($\mu_{\mathrm{Li}}$,
$\mu_{\mathrm{Ti}}$ and $\mu_{\mathrm{O}}$) were generated starting
from the assumption that \titanate\ can be formed from Li$_{2}$O
and TiO$_{2}$ via reaction \ref{eq:reaction} (this approach has
been adopted for simplicity, though we note that this is not the traditional
route for synthesis of \titanate\cite{Sinha:Li2TiO3_synthesis}),
\begin{equation}
\centering\mathrm{Li}_{2}\mathrm{O}_{\left(\mathrm{s}\right)}+
\mathrm{TiO}_{2\left(\mathrm{s}\right)}\rightarrow\mathrm{Li}_{2}
\mathrm{TiO}_{3\left(\mathrm{s}\right)}.\label{eq:reaction}
\end{equation}
The sum of the chemical potentials of the constituent species must
equal the total Gibbs free energy of the \titanate, i.e.
\begin{equation}
\centering\mu_{\mathrm{TiO}_{2}}\left(p_{O_{2}},T\right)+
\mu_{\mathrm{Li}_{2}\mathrm{O}}\left(p_{O_{2}},T\right)=
\mu_{\mathrm{Li}_{2}\mathrm{TiO}_{{3}\left(s\right)}}
\label{eq:chemical_potential1}
\end{equation}
 where, $\mu_{\mathrm{TiO}_{2}}\left(p_{\mathrm{O}_{2}},T\right)$
and $\mu_{\mathrm{Li}_{2}\mathrm{O}}\left(p_{\mathrm{O}_{2}},T\right)$
are the chemical potentials of Li$_{2}$O and TiO$_{2}$ within lithium
metatitanate as a function of oxygen partial pressure and temperature
and $\mu_{\mathrm{Li}_{2}\mathrm{TiO}_{{3}\left(s\right)}}$ is the
chemical potential of solid $\mathrm{Li}_{2}\mathrm{TiO}_{3}$. For
a solid $\mu\left(\mathrm{O}_{2}^{\circ},T^{\circ}\right)
\approx\mu\left(0,0\right)$ therefore the temperature and pressure
dependencies have been dropped. Two limiting cases are envisaged, one
in which the titanate is formed with excess Li$_{2}$O, ie.
$\mu_{\mathrm{Li}_{2}\mathrm{O}}\left(p_{\mathrm{O}_{2}},T\right)
=\mu_{\mathrm{Li}_{2}\mathrm{O}_{\left(s\right)}}$ and
$\mu_{\mathrm{TiO}_{2}}\left(p_{\mathrm{O}_{2}},T\right)=\mu_{\mathrm{Li}_{2}
\mathrm{TiO}_{{3}\left(s\right)}}-\mu_{\mathrm{Li}_{2}
\mathrm{O}_{\left(s\right)}}$ and similarly titania rich formation conditions
where $\mu_{\mathrm{TiO}_{2}}\left(p_{\mathrm{O}_{2}},T\right)=
\mu_{\mathrm{TiO}_{2\left(s\right)}}$. Here we assume Li$_{2}$O-rich
conditions, therefore $\mu_{\frac{1}{2}\mathrm{O_{2}}}
\left(p_{\mathrm{O}_{2}}^{\circ},T^{\circ}\right)$ can be determined from
the formation energy of Li$_{2}$O under standard conditions (taken from a
thermochemical database\cite{chase:janaftables}) and the DFT total energies
for Li$_{2}$O and lithium metal. $\mu_{\frac{1}{2}\mathrm{O_{2}}}
\left(p_{\mathrm{O}_{2}},T\right)$ can then be determined following
Finnis \etal\ \cite{finnis:oxygen_chemical_pot}, $\mu_{\mathrm{Li}}
\left(p_{\mathrm{O}_{2}},T\right)=1/2\left(
\mu_{\mathrm{Li}_{2}\mathrm{O}}^{DFT}-
\mu_{\frac{1}{2}\mathrm{O_{2}}}\left(p_{\mathrm{O}_{2}},T\right)\right)$
and similarly$\mu_{\mathrm{Ti}}\left(p_{\mathrm{O}_{2}},T\right)=
\mu_{\mathrm{TiO}_{2}}^{DFT}-2\mu_{\frac{1}{2}\mathrm{O_{2}}}
\left(p_{\mathrm{O}_{2}},T\right)$. For the purposes of this work a
temperature of 1000 K and oxygen partial pressure of 0.2 atm were selected.

In a periodic system the electrostatic energy is only finite if the
total charge on the repeat cell is zero. Therefore, when modelling
a charged defect in a simulation cell subject to PBCs, a uniform jellium
of charge is imagined to exactly neutralize the net charge on the
supercell\cite{dabo:electrostatics}. The electrostatic energy of a
periodically repeating finite system containing a point charge, $q$,
and a neutralizing background jellium is the Madelung energy,
\begin{equation}
\centering E=-\frac{q^{2}v_{M}}{2\epsilon}\label{eq:madelung}
\end{equation}
where $v_{M}$ is the Madelung potential (for cubic systems $v_{M}=\alpha/L$,
where $\alpha=2.8373$ and $L$ is the supercell size length). This
energy (scaled by $\epsilon$ to represent dielectric screening in
the material) arises due to the use of PBCs and is therefore an artifact
of the simulation technique and must be removed from the calculated
defect formation energy resulting in the charge correction proposed
by Leslie and Gillan\cite{leslie:coulombic_correction}:
\begin{equation}
\centering E_{f}^{\infty}=E_{f}(L)+\frac{q^{2}v_{M}}{2\epsilon}.
\label{eq:makov_payne}
\end{equation}
In highly ionic materials, defect charge distributions can be described
as point like, so this correction is adequate
\cite{komsa:charge_correction_discussion}, however when the defect charge
distribution is more diffuse the correction of Makov and
Payne\cite{makov:makov_payne_correction} is more appropriate.
Castleton and co-workers proposed an extrapolation
procedure\cite{castelton:defects_InP,castelton:defects_convergence_InP}
for the study of defects in InP whereby the limit of a fit to a series
of formation energies obtained from supercells of increasing size was
used to determine the formation energy of the isolated defect. However,
uniformly scaling all axes simultaneously rapidly increases the number
of atoms in the supercell. Consequently it is only computationally feasible
to sample the smallest multiplications resulting in too few points to allow
a reliable fit to the data. Hine \etal\ thus suggested an improvement to this
scheme based on simulation of supercells comprising different multiples of the
primitive cell along different axes\cite{hine:coulombic_correction},
with $v_{M}$ calculated separately for each cell. By plotting the
defect formation energy as a function of $v_{M}$ and fitting a function
of the form $E_{f}\left(v_{M}\right)=E_{f}^{\infty}+bv_{M}$, it is
possible to determine the formation energy of the defect in the dilute
limit from the intercept with the $y$-axis, and an effective permittivity
can be extracted from the gradient as $b=-q^{2}/2\epsilon$.

The constant $v_{M}$ can be found using Ewald summation \cite{ewald:ewald_sum}:
\begin{equation}\label{eq:madelung2}
v_{M} =
 \sum_{\mathbf{R}_{i}}^{i\not=0}\frac{\text{erfc}
 \left(\gamma\sqrt{|\mathbf{R}_{i}|}\right)}{|\mathbf{R}_{i}|}+
 \sum_{\mathbf{G}_{i}}^{i\not=0}\frac{4\pi}{V_{c}}\frac{\exp
 \left(-\mathbf{G}_{i}^{2}/4\gamma^{2}\right)}{\mathbf{G}_{i}^{2}}
-\frac{2\gamma}{\sqrt{\pi}}-\frac{\pi}{V_{c}\gamma^{2}}.
\end{equation}
where the sum extends over all vectors of the direct ($\mathbf{R}_{i}$)
and reciprocal ($\mathbf{G}_{i}$) lattices, $\gamma$ is a suitably chosen
convergence parameter and $V_c$ is the volume of the supercell. $v_{M}$
is normally positive and hence the Madelung energy is normally negative
as it is dominated by the interactions of the point charge and the
canceling background jellium which is on average closer than the periodic
images. For long, thin supercells this is no longer the case as the
electrostatics are now dominated by the interactions of neighbouring
point charges and so $v_{M}$ is negative. This can, however,
be viewed as an advantage since simulations can be performed on supercells
where $v_{M}$ is both negative and positive and the results interpolated
to $v_{M}=0$ rather than performing an extrapolation outside the
range for which data is available.

\begin{figure}[P]
\clearpage
\centering{}\includegraphics[width=0.7\columnwidth]{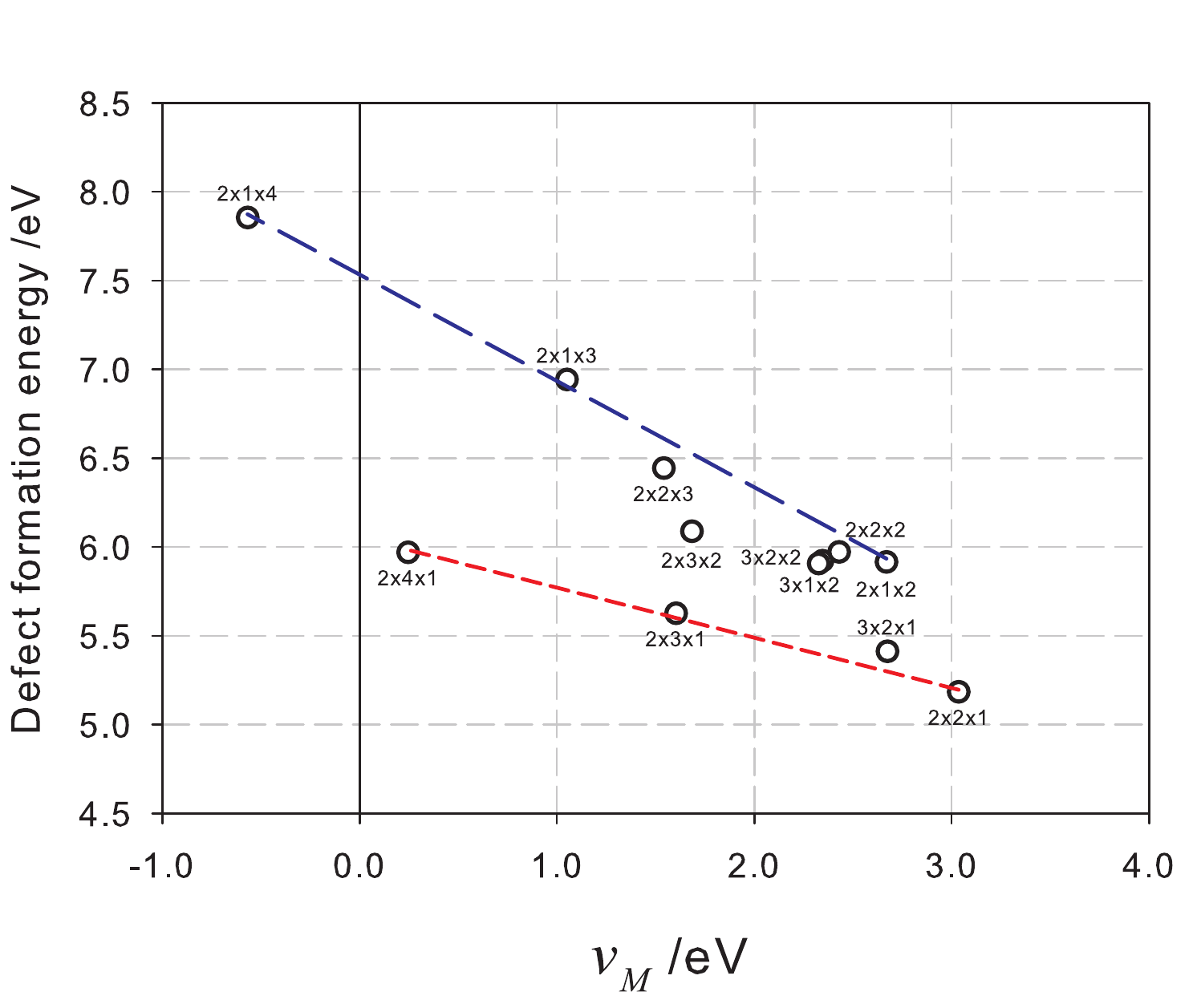}
\caption{\label{figure:uncorrected}Formation energy of the
V$_{\mathrm{Ti}}^{-4}$ defect for a range of supercells with differing
$v_{M}$. The wide variation in the points demonstrates that is not
possible to fit a single straight line of the
$E_{f}\left(v_{M}\right)=E_{f}^{\infty}+bv_{M}$ to the data.
}\newpage
~
\newpage
\end{figure}

Fig.~\ref{figure:uncorrected} shows the formation energy of the
V$_{\mathrm{Ti}}^{-4}$ defect as a function of $v_{M}$
for a range of supercell shapes and sizes. The data display a wide
variation and it is not possible to extract a single value for
$E_{f}^{\infty}$. The origin of this variation may be deduced by
examining subsets of the data. Shown in Fig.~\ref{figure:uncorrected} are fits
of the form $E_{f}\left(v_{M}\right)=E_{f}^{\infty}+bv_{M}$ to defect
formation energies calculated in supercells created by extrapolating
in the number of repeat units along the $b$- and $c$-axes independently,
i.e. 2$\times{m}\times$1 (for $m$= 2,3 and 4) and 2$\times$1$\times{n}$
(for $n$ = 2,3 and 4). As the effective permittivity can be related to the
gradient of such a fit it is apparent that there is a different level of
charge screening present along these crystallographic axes. The effective
dielectric constant along $b$ (28.3) is predicted to be more than
double that along $c$ (13.4).

To account for anisotropy in the screening, the dielectric constant
in Eq. \ref{eq:makov_payne} must be replaced by a
tensor\cite{rurali:dielectric_nanowires}, denoted $\bar{\epsilon}$.
For a monoclinic crystal such as \titanate\ the dielectric tensor has four
non-zero components, as shown below\cite{nye:crystals}:
\begin{equation}
\bar{\epsilon}=\left[\begin{array}{ccc}
\epsilon_{11} & 0 & \epsilon_{13}\\
0 & \epsilon_{22} & 0\\
\epsilon_{13} & 0 & \epsilon_{33}
\end{array}\right]\,.\label{eq:tensor}
\end{equation}
This tensor can then be incorporated into the Ewald summation to give
a \emph{screened} Madelung potential, $v_{M}^{\mathrm{scr}}$, in
the general case
\cite{fischerauer:anisotropic_medium,rurali:dielectric_nanowires}:
\begin{equation}\label{eq:modified_madelung}
v_{M}^{\mathrm{scr}}  = \sum_{\mathbf{R}_{i}}^{i\not=0}
\frac{1}{\sqrt{\det{\bar{\epsilon}}}}
\frac{\text{erfc}\left(\gamma\sqrt{\mathbf{R}_{i}\cdot
\bar{\epsilon}^{-1}{\cdot}\mathbf{R}_{i}}\right)}{\sqrt{\mathbf{R}_{i}{\cdot}
\bar{\epsilon}^{-1}{\cdot}\mathbf{R}_{i}}}+\sum_{\mathbf{G}_{i}}^{i\not=0}
\frac{4\pi}{V_{c}}\frac{\exp\left(-\mathbf{G}_{i}{\cdot}\bar{\epsilon}{\cdot}
\mathbf{G}_{i}/4\gamma^{2}\right)}{\mathbf{G}_{i}{\cdot}\bar{\epsilon}{\cdot}
\mathbf{G}_{i}}-\frac{2\gamma}{\sqrt{\pi\det{\bar{\epsilon}}}}-\frac{\pi}{V_{c}
\gamma^{2}}.
\end{equation}

Eq.~\ref{eq:modified_madelung} implies that it is necessary to determine
the dielectric tensor for each defect cell, which while possible
using Density Functional Perturbation Theory (DFPT)\cite{refson:DFPT},
is computationally prohibitive. % particulary for larger supercells.
Here we investigate two possible methods that circumvent this problem.

In the region far from the defect the atomic positions (and consequently
the dielectric properties) will remain largely unaffected by the presence
of the defect, however, in the region immediately surrounding the defect
the screening properties may be strongly perturbed. If the perturbed region
is small relative to the simulation supercell then the dielectric properties
of the whole cell may not undergo a substantial modification, As a first
approximation, we therefore try applying the dielectric tensor for the perfect
\titanate\ crystal to the all defective systems.

In our second approach a function
$E_{f}(v_{M})=-(q^{2}/2)v_{M}+E_{f}^{\infty}$ is fitted to the defect
formation energies determined for a number of different cell shapes and
sizes. This fitting procedure is slightly unusual as it is the values in
the $x$-axis that are modified by optimising the elements of
$\bar{\epsilon}^{\mathrm{eff}}$. Optimised values of $E_{f}^{\infty}$ and
the associated elements of $\bar{\epsilon}^{\mathrm{eff}}$ were obtained
using a Nelder-Mead simplex algorithm\cite{nelder:algorithm}.

\section{Results and discussion}

The dielectric tensor for \titanate\ was calculated using DFPT and the
norm-conserving pseudo-potentials. The results, presented in Table
\ref{table:results}; show that there is indeed a significant level of
anisotropy in the dielectric tensor. Examining only the principal (diagonal)
elements of $\bar{\epsilon}^{\mathrm{DFPT}}$ we can see that the magnitude
of $\epsilon_{33}^{\mathrm{eff}}$ is less than half that of
$\epsilon_{11}^{\mathrm{eff}}$ and $\epsilon_{22}^{\mathrm{eff}}$. Taking the
tensor average gives a value of 30.5, which can be compared to a value of 24
for a polycrystalline \titanate\ sample\cite{bian:li2tio3_dielectric} (this
value has been corrected to represent the theoretical density). The
discrepancy between the experimental and theoretical dielectric properties may
arise due to the inherent inability of DFT simulations to accurately reproduce
experimentally observed band gaps. The values also deviate from those
predicted by examining the subsets of the uncorrected data determined
from Fig. \ref{figure:uncorrected}.

\begin{figure}[P]
\centering{}\includegraphics[width=0.7\columnwidth]{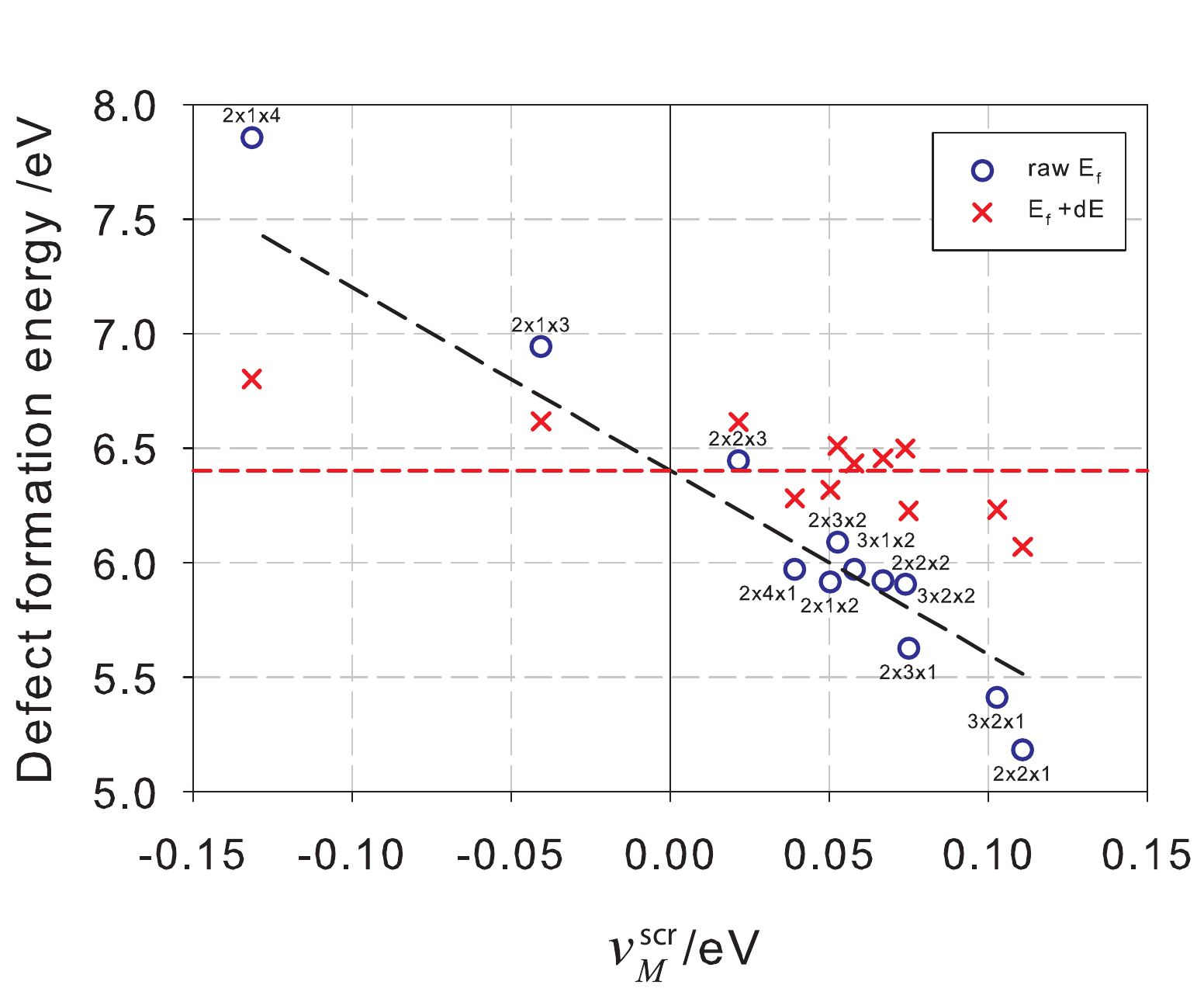}
\caption{\label{figure:DFPT}Formation energy of V$_{\mathrm{Ti}}^{-4}$ as
a function of $v_{M}^{\mathrm{scr}}$, where $\bar{\epsilon}^{\mathrm{DFPT}}$
has been used in the calculation of $v_{M}^{\mathrm{scr}}$. The black dashed
line represents a fit of the form given in Eq~\ref{eq:makov_payne} to the raw
data and the red dashed line is fitted to the corrected data.}
\end{figure}

\begin{figure}[P]
\clearpage
\centering{}\includegraphics[width=0.7\columnwidth,trim=0cm 3.5cm 0cm 0cm]{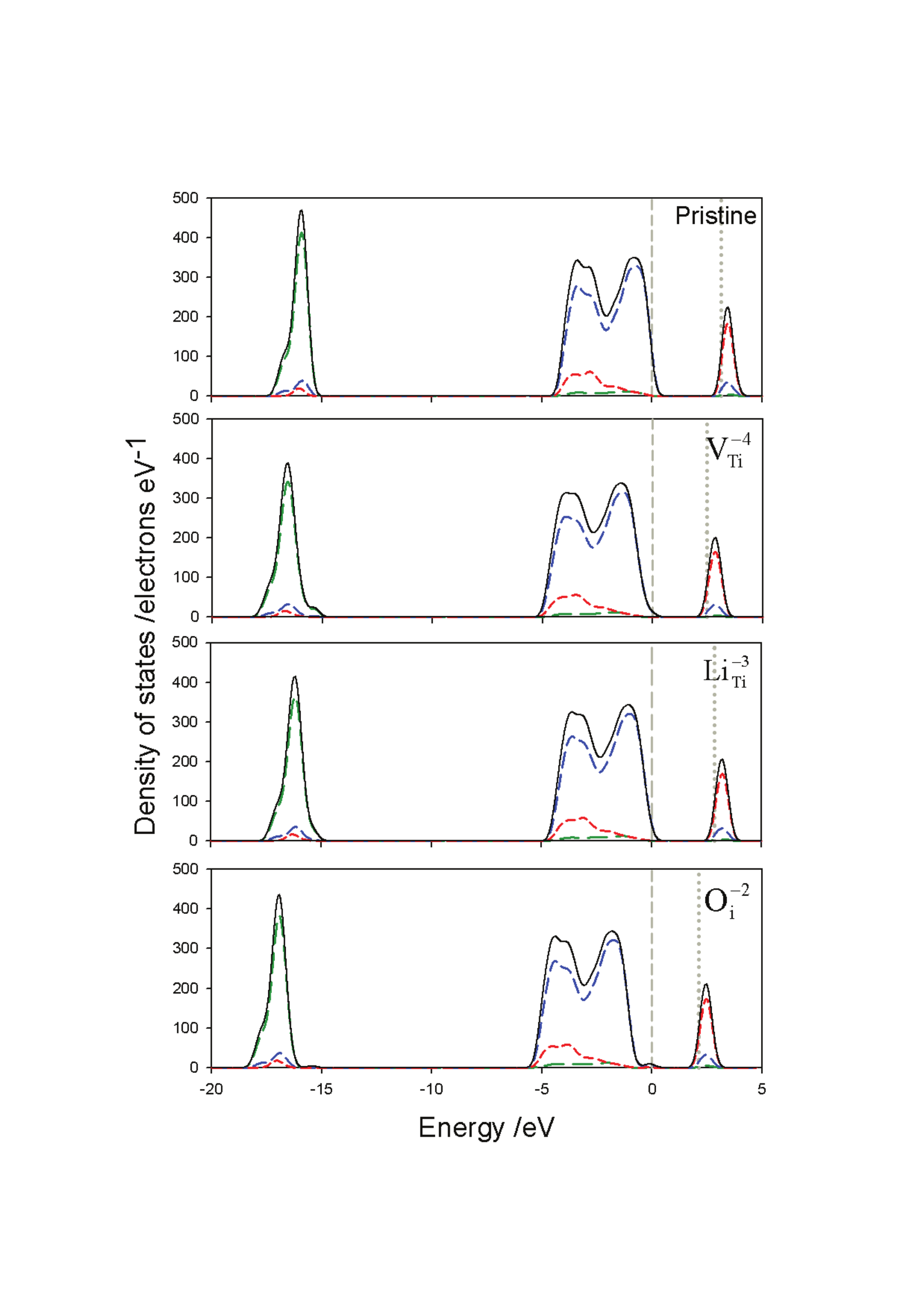}
\caption{\label{figure:dos}Densities of states for perfect \titanate\ and the
V$_{\mathrm{Ti}}^{-4}$, Li$_{\mathrm{Ti}}^{-3}$ and O$_i^{-2}$ defects. The
long dashed green lines, intermediate dashed blue lines and short dashed red
lines correspond to $s$-, $p$- and $d$-derived states respectively, with the
sum plotted using the solid black line. The valence band maximum is
indicated with a dashed grey vertical line and the conduction band minimum
with a dotted grey vertical line.
}\newpage
~
\newpage
\end{figure}

Corrections such as that of Makov-Payne\cite{makov:makov_payne_correction}
are often performed with $\epsilon$ obtained from either DFPT or experiment.
Fig. \ref{figure:DFPT} shows defect formation energies for
V$_{\mathrm{Ti}}^{-4}$ for a selection of supercells as a function of
$v_{M}^{\mathrm{scr}}$, employing $\bar{\epsilon}^{\mathrm{DFPT}}$.
The data points show that while the level of scatter present in Fig.
\ref{figure:uncorrected} has been reduced there is also a poor adherence
to the linear relationship with gradient $-q^2/2$ expected from
Eq.~\ref{eq:makov_payne}. This discrepancy arises as the use of the dielectric
tensor calculated for the perfect cell, which thus neglects the atomic
relaxations and the consequent modification of the local screening in the
vicinity of the defect. The modification of the dielectric properties of the
supercell is further supported by the change in the band gap of the material
upon introduction of the defect. Plotted in Fig. \ref{figure:dos} are the
Densities of States (DOS) for the perfect \titanate\ as well as the defect
containing supercells (all DOS are produced for the 2$\times2\times2$ supercell).
Fig. \ref{figure:dos} shows that the bandgaps in the defect containing
supercells are reduced relative to the perfect supercell
($E_g(\mathrm{V}_{\mathrm{Ti}}^{-4}) = 2.46$ eV,
$E_g(\mathrm{Li}_{\mathrm{Ti}}^{-3}) = 2.86$ eV and
$E_g(\mathrm{O}_{i}^{-2}) = 2.12$ eV),  which would suggest a perceptible
change in the dielectric properties of the cell.

\begin{sidewaysfigure}[P]
%\begin{figure*}
\subfigure[\label{figure:Ti_vac1-4}]{
\includegraphics[width=0.30\columnwidth]{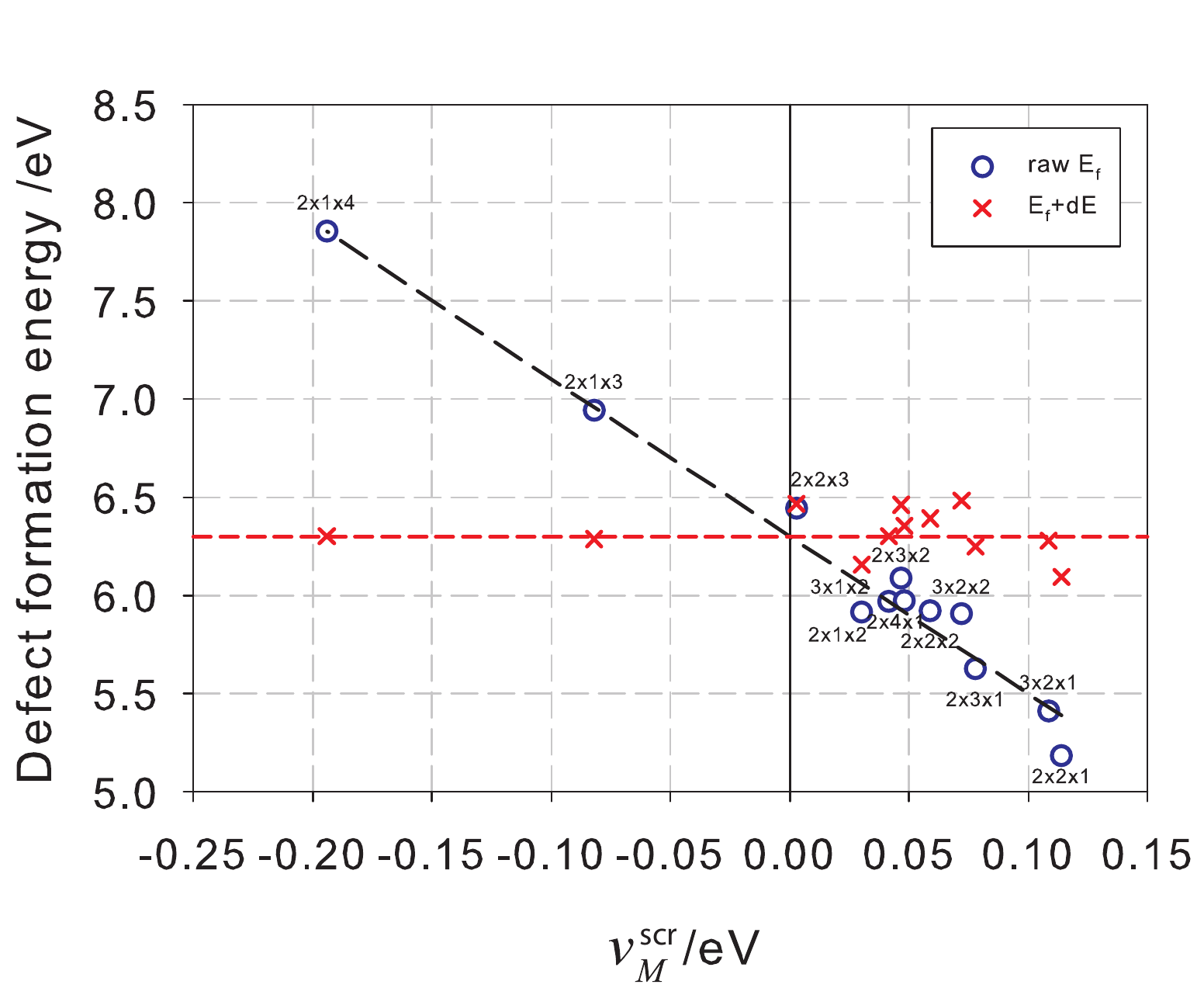}}
\subfigure[\label{figure:Liti1-3}]{
\includegraphics[width=0.30\columnwidth]{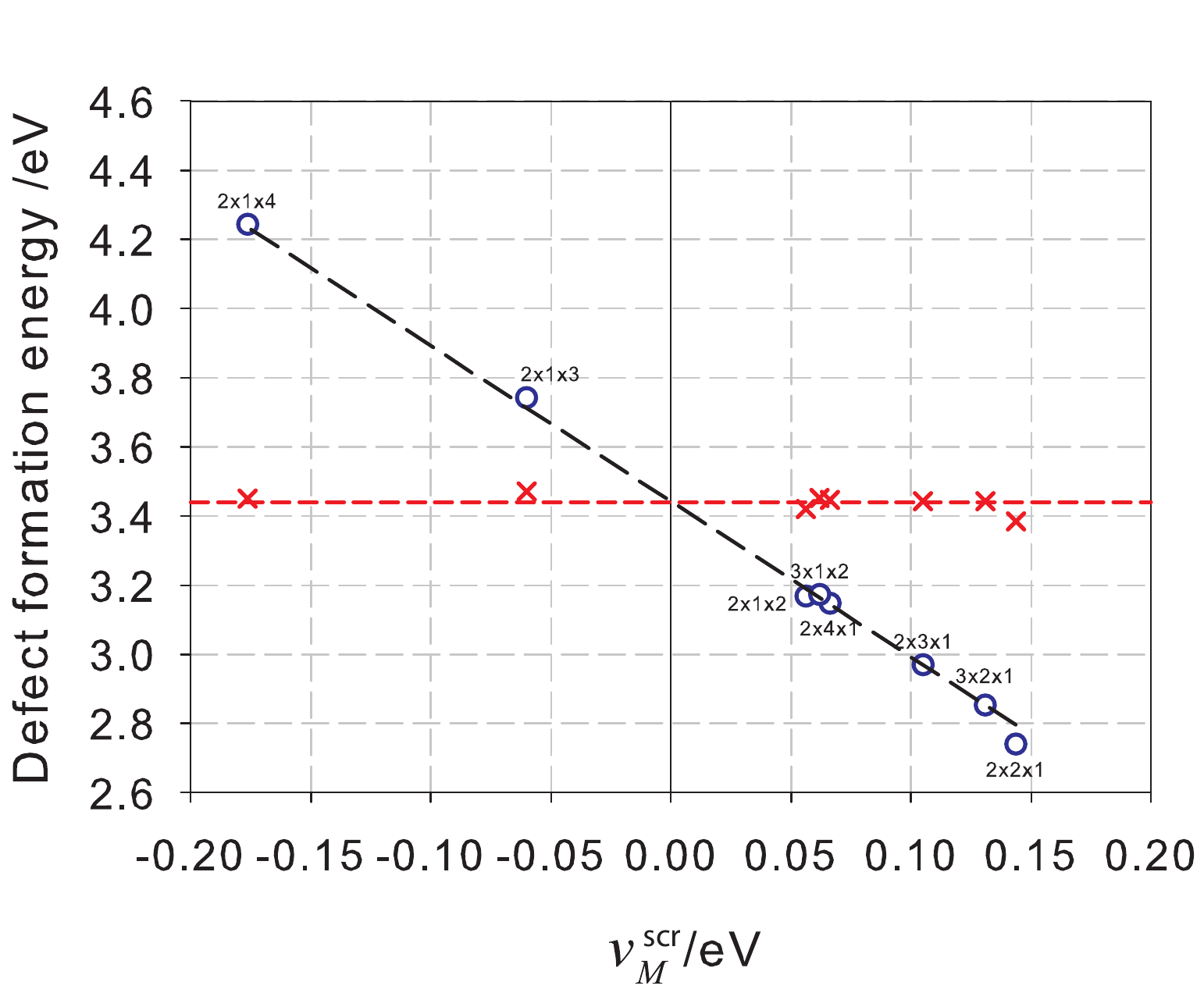}}
\subfigure[\label{figure:O_int1-2}]{
\includegraphics[width=0.30\columnwidth]{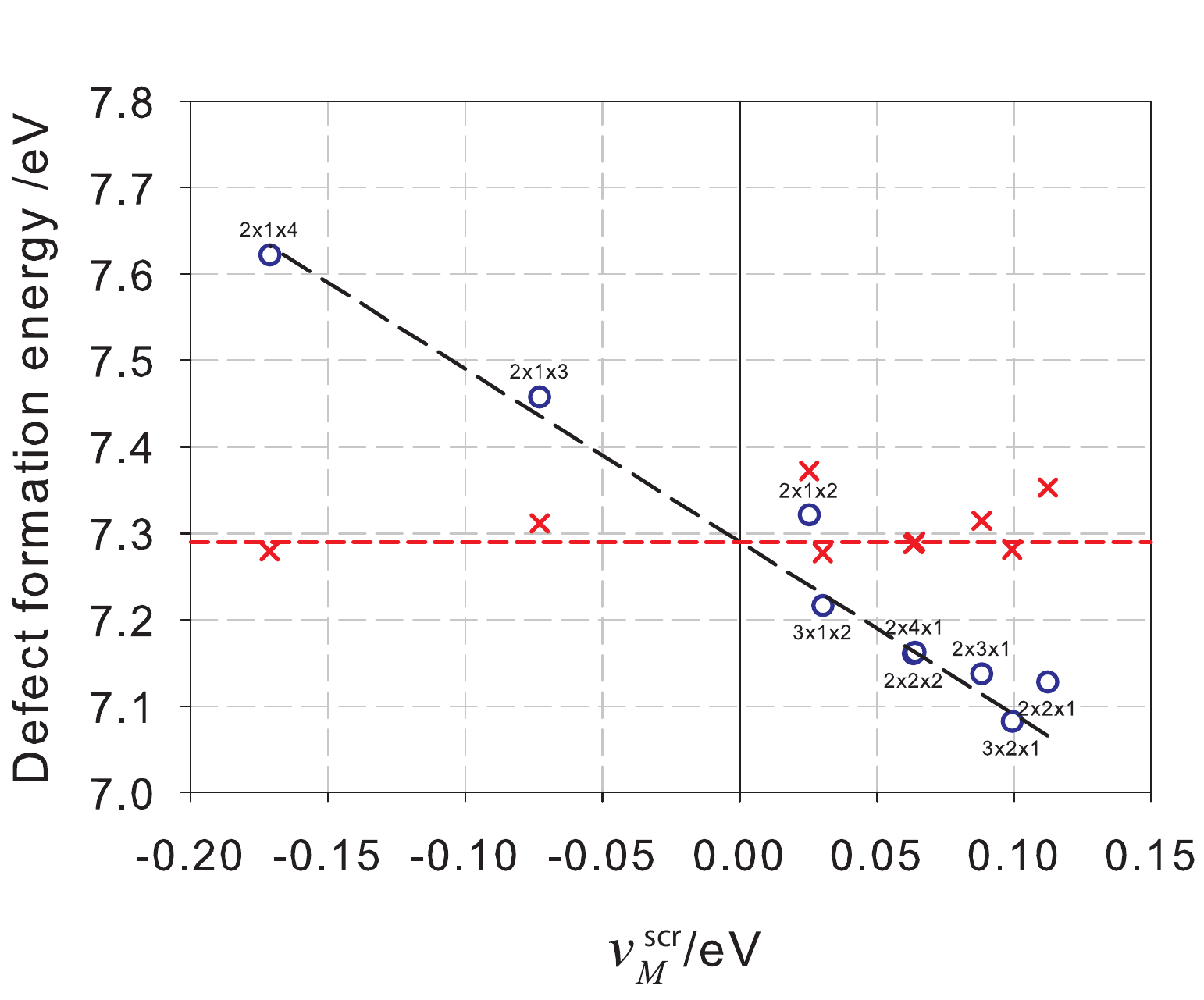}}
\caption{\label{figure:plots}Defect formation energy as a function
of $v_{M}^{\mathrm{scr}}$ for the (a) V$_{\mathrm{Ti}}^{-4}$, (b)
Li$_{\mathrm{Ti}}^{-3}$ and (c) O$_{i}^{-2}$ defects, where nonzero
elements of $\bar{\epsilon}^{\mathrm{eff}}$ are fitted to the data.
The wide variation in the data shown in Fig.~\ref{figure:uncorrected}
has disappeared, so one can interpolate to $v_{M}^{\mathrm{scr}}=0$
and extract defect formation energies in the dilute limit. Note that
simulations in the largest 3$\times$2$\times$2, 2$\times$3$\times$2
and 2$\times$2$\times$3 supercells were only performed for the
V$_{\mathrm{Ti}}^{-4}$
defect.}
%\end{figure*}
\end{sidewaysfigure}
In order to incorporate the change in the dielectric properties of
the supercells induced by the defect, we instead fit the elements
of $\bar{\epsilon}$ to give $\bar{\epsilon}^{\mathrm{eff}}$.
$\bar{\epsilon}^{\mathrm{eff}}$ is then effectively an averaged picture
of the dielectric tensors for the supercells included in the fit.
Presented in Fig.~\ref{figure:plots} are plots of the
formation energies as a function of $v_{M}^{\mathrm{scr}}$ after
the fitting procedure has been performed for the V$_{\mathrm{Ti}}^{-4}$,
Li$_{\mathrm{Ti}}^{-3}$ and O$_{i}^{-2}$ defects. The three plots
in Fig.~\ref{figure:plots} show that by fitting the elements of
$\bar{\epsilon}$ it is possible to substantially improve the correlation
between the data and the relationship given in Eq.~\ref{eq:makov_payne},
thus allowing a single linear function to be fitted to the data and a dilute
limit defect formation energy to be extracted. Residual errors associated
with the fitting process are around 0.1 eV and likely arise either
from dipole-dipole or monopole-quadrupole interactions not accounted
in Eq.~\ref{eq:makov_payne}, or from changes in atomic configurations
for cells with one small value of $l$,$m$,$n$. The relatively small
errors justify our treatment of the defect charge state as point-like.
However in complex systems where the defect charge is less localised,
or for defect clusters, this approximation may no longer hold.

\begin{table}
\caption{\label{table:results}Effective permittivity tensor
$\bar{\epsilon}^{\mathrm{eff}}$, and dilute limit defect formation energies
$E_{f}^{\infty}$, for several defect species.}
\centering{}%
\begin{tabular}{|c|c|c|c|c|c|}
\hline
Species & $\epsilon_{11}^{\mathrm{eff}}$  & $\epsilon_{22}^{\mathrm{eff}}$  &
$\epsilon_{33}^{\mathrm{eff}}$  & $\epsilon_{13}^{\mathrm{eff}}$  &
$E_{f}^{\infty}$ /eV \tabularnewline
\hline
\titanate\ (DFPT)  & 36.1  & 37.8  & 17.8  & -5.0  & - \tabularnewline
\hline
$\mathrm{V}_{\mathrm{Ti}}^{-4}$  & 37.4  & 37.4  & 14.4  & -1.0  & 6.3
\tabularnewline
$\mathrm{Li}_{\mathrm{Ti}}^{-3}$  & 34.9  & 35.0  & 13.9  & -12.1  & 3.5
\tabularnewline
O$_{i}^{-2}$  & 33.7  & 55.6  & 16.5  & -8.9  & 7.3 \tabularnewline
\hline
\end{tabular}
\end{table}

\begin{figure}[P]
\centering \subfigure[\label{figure:perfect}]{
\includegraphics[width=0.33\columnwidth]{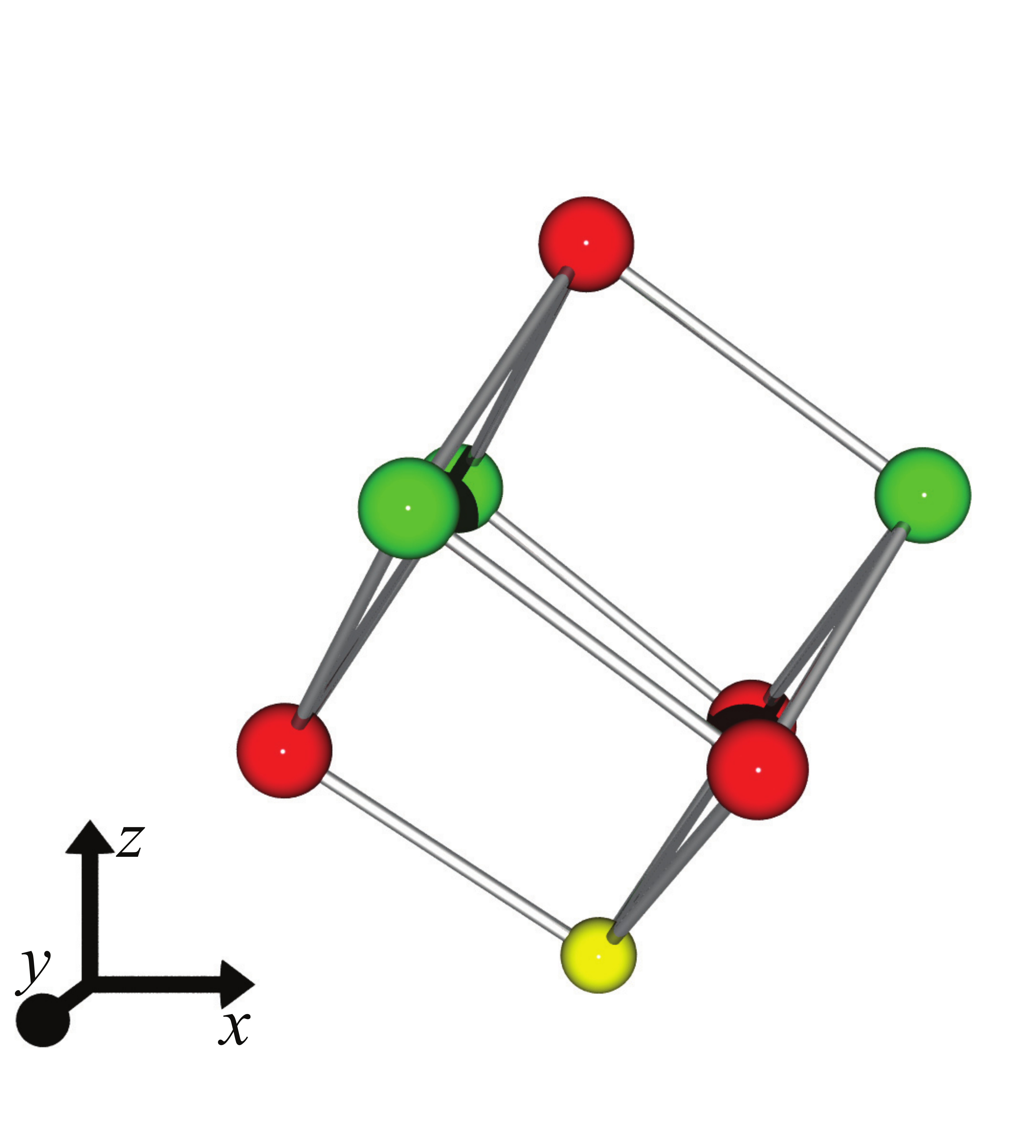}}
\subfigure[\label{figure:output}]{
\includegraphics[width=0.33\columnwidth]{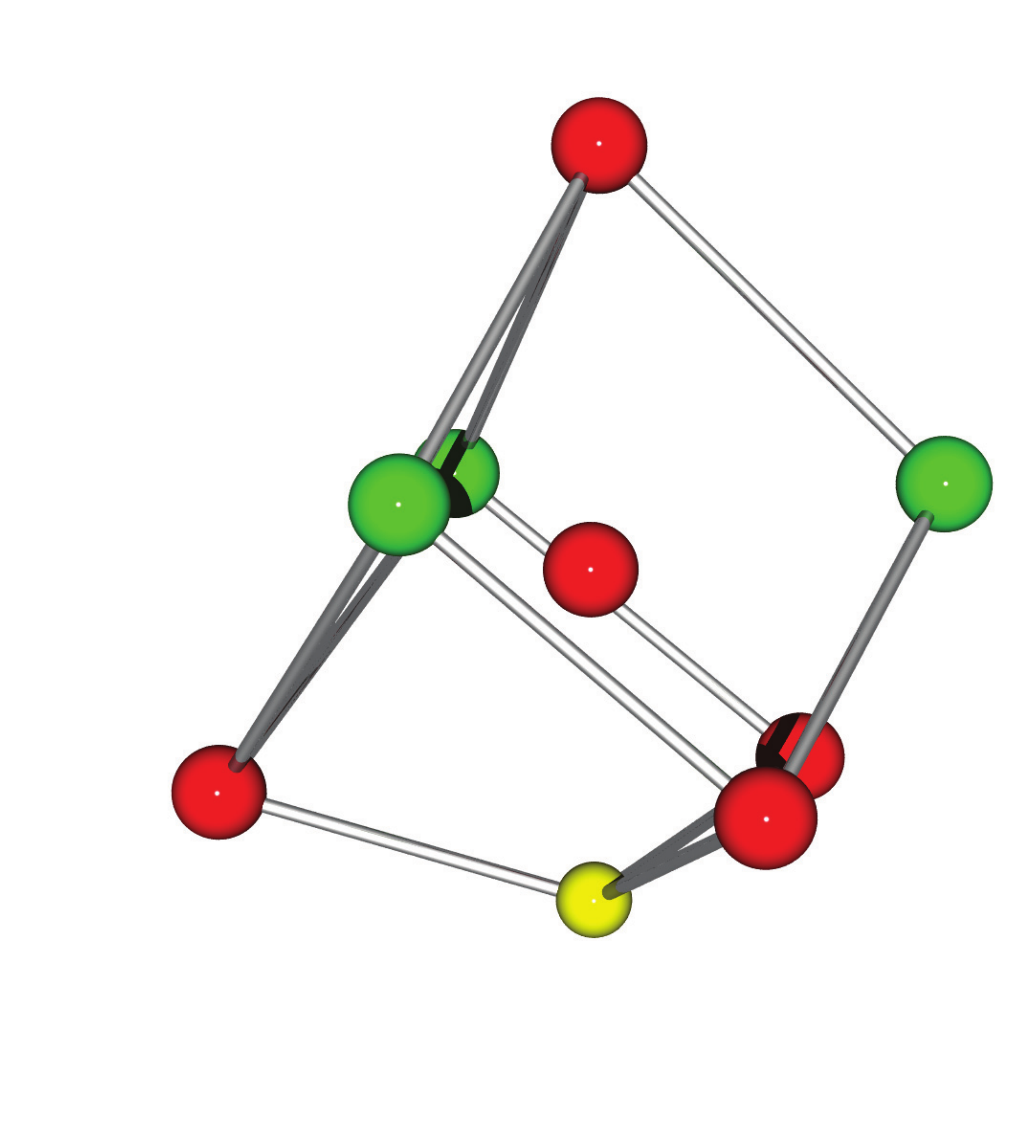}}
\caption{\label{figure:relaxation}(a) Atomic arrangement surrounding an as
yet unoccupied interstitial site in \titanate\ and (b) same atoms
after introduction of an O$_{i}^{-2}$ defect. Clearly visible in
(a) is the distorted rocksalt structure of \titanate\ and the distortion
arising arising due to the defect is illustrated in (b). It is this
distortion that causes the significant change in
$\bar{\epsilon}^{\mathrm{eff}}$ for this defect. All atom positions
have been extracted from simulations in the 2$\times$2$\times$2
supercells.}\clearpage
\end{figure}

The fitted elements of $\bar{\epsilon}^{\mathrm{eff}}$ and the resulting
dilute-limit defect formation energies are presented in
Table \ref{table:results}. In the case of the V$_{\mathrm{Ti}}^{-4}$ and
Li$_{\mathrm{Ti}}^{-3}$ defects the degree of atomic relaxation is relatively
small and the concomitant differences between $\bar{\epsilon}^{\mathrm{DFPT}}$
and $\bar{\epsilon}^{\mathrm{eff}}$ are also modest. The level of
local distortion resulting from the introduction of an O$_{i}^{-2}$
defect is much greater than for the other defects as depicted in
Fig. \ref{figure:relaxation}. Furthermore the reduction in the bandgap is
greatest for this defect which is also consistent with it displaying
the most significant perturbation in its dielectric properties. It is
this higher level of relaxation that leads to the increased difference
between $\bar{\epsilon}^{\mathrm{DFPT}}$ and $\bar{\epsilon}^{\mathrm{eff}}$.
At larger system sizes we would expect to recover values of
$\bar{\epsilon}^{\mathrm{eff}}$  increasingly close to those
of the perfect crystal, but the current results indicate that at the fairly
small system sizes obtainable here, it is appropriate to fit to the observed
results.

\section{Conclusions}

In summary, we have proposed an extension of the Madelung extrapolation
procedure\cite{hine:coulombic_correction} for the calculation of defect
formation energies in the dilute limit. This is achieved by incorporating
the effect of charge screening, via the dielectric tensor, into the
calculation of the Madelung potential (via Eq.~\ref{eq:modified_madelung})
and fitting the elements of the tensor and the desired dilute-limit formation
energy to defect formation energies calculated in a range of supercells. We
have applied the method to \titanate, which has a monoclinic structure and a
highly anisotropic dielectric tensor, and demonstrated its ability to
determine defect formation energies converged to within around 0.1 eV even for
such systems. In principle this method is applicable to systems of
any shape and dielectric properties. Even in cubic supercells, local
relaxation, such as that arising from defect clusters, may break the
crystal symmetry such that the dielectric properties are anisotropic,
necessitating a tensor representation of dielectric properties. We have
also further highlighted the importance of incorporating the effect of
lattice relaxation on the dielectric properties of the material when applying
a finite-size correction method based on the
Makov-Payne\cite{makov:makov_payne_correction} approximation.

\section{Acknowledgments}

Computational resources were provided by the Imperial College High
Performance Computing Centre. Prof. Robin Grimes is thanked for useful
discussions. NDMH acknowledges the support of EPSRC Grants EP/G05567X/1
and EP/J015059/1, and the Leverhulme Trust.

%\bibliography{papers}
%merlin.mbs 2010-03-15 4.21a (PWD, AO, DPC)
%Control: key (0)
%Control: author (8) initials jnrlst
%Control: editor formatted (1) identically to author
%Control: production of article title (-1) disabled
%Control: page (0) single
%Control: year (1) truncated
%Control: production of eprint (0) enabled
%

\end{document}